\documentclass[runningheads]{llncs}
\usepackage{graphicx}
\usepackage{fancyhdr}
\usepackage{amsmath,amssymb,amsfonts}
\usepackage{algorithm, algorithmic}
\usepackage{epstopdf}
\usepackage{cite}
\usepackage{enumitem}
\usepackage{array}

\begin{document}
	\title{BiSample: Bidirectional Sampling for Handling Missing Data with Local Differential Privacy}
	
	\author{Lin Sun\textsuperscript{1,2}, Xiaojun Ye\textsuperscript{1}, Jun Zhao\textsuperscript{2}, Chenhui Lu\textsuperscript{1}, and Mengmeng Yang\textsuperscript{2}}
	\authorrunning{~}

	\institute{\textsuperscript{1}{Key Laboratory for Information System Security, School of Software, Tsinghua University, Beijing, China}\\ \textsuperscript{2}{School of Computer Science and Engineering, Nanyang Technological University, Singapore}\\ dr.forestneo@gmail.com, yexj@mail.tsinghua.edu.cn, luch18@mails.tsinghua.edu.cn, JunZhao@ntu.edu.sg, melody.yang@ntu.edu.sg}
	\maketitle              

\thispagestyle{fancy}
\pagestyle{fancy}
\lhead{This paper appears as a full paper in the Proceedings of 25th International Conference on Database Systems for Advanced Applications (DASFAA 2020).\\~}
\cfoot{\thepage}
\renewcommand{\headrulewidth}{0pt}
\renewcommand{\footrulewidth}{0pt}

	\begin{abstract}
		Local differential privacy (LDP) has received much interest recently. In existing protocols with LDP guarantees, a user encodes and perturbs his data locally before sharing it to the aggregator. In common practice, however, users would prefer not to answer all the questions due to different privacy-preserving preferences for different questions, which leads to data missing or the loss of data quality. In this paper, we demonstrate a new approach for addressing the challenges of data perturbation with consideration of users' privacy preferences. Specifically, we first propose BiSample: a bidirectional sampling technique value perturbation in the framework of LDP. Then we combine the BiSample mechanism with users' privacy preferences for missing data perturbation. Theoretical analysis and experiments on a set of datasets confirm the effectiveness of the proposed mechanisms.
		
		
		

	\end{abstract}
	
	\section{Introduction}
	
	With the development of big data technologies, numerous data from users' side are routinely collected and analyzed. In online-investigation systems, statistical information, especially the frequency and mean values can help investigators know about the investigated population. However, users' data are collected at the risk of privacy leakage. Recently, local differential privacy (LDP)~\cite{duchi2013local} has been proposed as a solution to privacy-preserving data collection and analysis since it provides provable privacy protection, regardless of adversaries' background knowledge. Usually, protocols with LDP guarantees can be broken down into an Encode-Perturb-Aggregate paradigm. For single round of data sharing, each user \textbf{encodes} his value (or tuple) into a specific data format, and then \textbf{perturbs} the encoded value for privacy concerns. At last, all the perturbed data are \textbf{aggregated} by an untrusted collector. Mechanisms with LDP guarantees have been implemented in many real-world data collecting systems, such as Google's RAPPOR~\cite{erlingsson2014rappor} and Microsoft's telemetry data analyzing system~\cite{ding2017collecting}.

	Although the LDP can balance the users' privacy and data utilities, existing solutions assume that the investigated users follow the collecting process truthfully. However, in an investigation system, even though the investigator claims the collection process satisfies LDP, individuals may refuse to confide some specific questions due to following considerations: 1) the provided privacy-preserving level is not as expected, or 2) he just doesn't want to tell anything about the question. For example, if an investigator designed a $\ln 3$-LDP mechanism for personal-related data analyzing, those who think the privacy-preserving is good enough would provide the real value for perturbation, while those who extremely care about their healthy states might evade certain questions such as ``Do you have cancer?'' because they think the in-built privacy-preserving guarantee by LDP mechanism is not private enough. As existing perturbation solutions demand an input, these individuals would randomly pick an answer (or just answer ``No'') and use it for perturbation (we call these \textbf{fake answers}). In the perturbed space, the fake answers will lead to evasive bias.
	
	\begin{figure}[htbp]
		\centering
		\includegraphics[width=120mm]{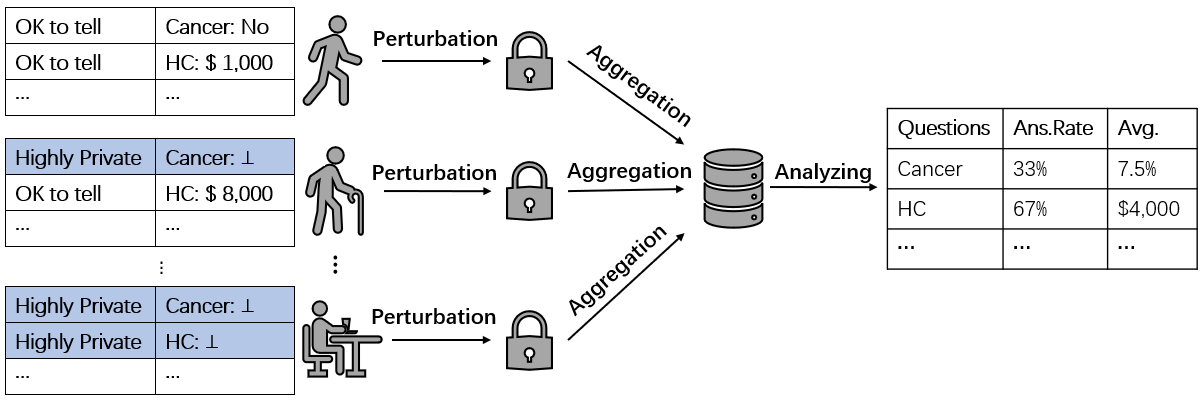}
		\caption{Missing Data Collecting and Analyzing Framework. For space consideration, we use HC to represent the Hospitalization Cost and use Ans.Rate to represent the rate of individuals who use real value for perturbation. The symbol $\bot$ occurs when users are not willing to provide real value, even in the LDP framework.}
		\label{fig: system framework}
	\end{figure}
	
	In this paper, we consider a ``providing null-value'' procedure to avoid the fake answers when users perturb their data. Fig.~\ref{fig: system framework} lists users' privacy preferences and original answers of each investigated question. Instead of sending a fake answer when the provided privacy-preserving level is not as expected, the investigated user sends a null-value to avoid a biased estimated result. The untrusted aggregator wants to analyze basic statistical information, for each question, 1) how many people provide the real value, 2) what is the frequency/mean of the whole investigated population.
	
	We for the first time consider the influence of users' cooperation on the estimation accuracy. We first propose a bi-directional sampling mechanism called BiSample and use it for numerical value perturbation. Then we extend the BiSample to be capable of the null-value perturbation while still being locally differentially private. In general, this paper presents following contributions:
	
	\begin{itemize}
		\item For the first time, we consider that not all users will provide the true data for perturbation in a collecting and analyzing framework. Our proposed missing data perturbation framework provides new insights into improving data utilities by modeling users' privacy-preserving preferences.
		\item We propose BiSample, a bi-directional sampling mechanism for data perturbation. Literately, the BiSample mechanism can replace the Harmony~\cite{nguyen2016collecting} solution for mean estimation. Furthermore, we extend the BiSample to be capable of perturbing null-value data. Our mechanism allows to analyze the rate of users who provided true answers and can be used for frequency/mean estimation with LDP guarantees.
		\item As the proposed framework can be used to estimate the rate of users who provide real value under privacy budget $\epsilon$, the BiSample mechanism can be used to study how to set privacy budget appropriately by the aggregator.
		\item Experimentally, the proposed mechanism achieves lower estimation error than existing mechanisms. 
	\end{itemize}
	
	This paper is organized as follows. In Section~\ref{sec: preliminaries}, we provide the necessary background of LDP and define the problem of analyzing missing data in our framework. Then we propose the BiSample mechanism in Section~\ref{sec: bisample} and apply BiSample for missing data in Section~\ref{sec: bisample-md}. The evaluations of the proposed mechanism are shown in Section~\ref{sec: experiment}. At last, the whole paper is concluded in Section~\ref{sec: conclusion}.
	
	\section{Preliminaries and Problem Definition}
	\label{sec: preliminaries}
	\subsection{Local Differential Privacy (LDP)}
	
	\begin{definition}[Local Differential Privacy~\cite{duchi2013local,bassily2015local,duchi2014privacy}]
		A randomized mechanism $\mathcal{M}(\cdot)$ achieves $\epsilon$-local differential privacy if and only if for every two input tuples $t_1, t_2$ in the domain of $\mathcal{M}$, and for any output $t^* \in \operatorname{Range}(\mathcal{M})$ that:
		\begin{equation}
		\Pr[\mathcal{M}(t_1)=t^*] \le \exp(\epsilon) \cdot \Pr[\mathcal{M}(t_2)=t^*]
		\end{equation}
	\end{definition}
	
	Unlike earlier attempts to preserve privacy, such as $k$-anonymity~\cite{samarati1998protecting} and $l$-diversity~\cite{li2007t}, the LDP retains ``plausible deniability'' of sensitive information. The LDP has been used in a variety of application areas since proposed, such as heavy hitters estimation~\cite{bassily2017practical,qin2016heavy,bassily2015local}, histogram estimation~\cite{wang2016private,cormode2018marginal}, and marginal release~\cite{cormode2018marginal}.
	
	The canonical solution towards LDP is the randomized response (RR~\cite{Warner1965,dwork2014algorithmic}). Specifically, to collect sensitive information from users, e.g., whether the patient is a HIV carrier, RR is used for perturbing the actual answers while still guarantees that i) each user's answer provides plausible deniability, ii) the aggregator can get an unbiased estimation over the whole population. Many start-of-the-art mechanisms use RR as a core part to provide privacy guarantees, such as the LDPMiner~\cite{qin2016heavy}, LoPub~\cite{ren2018lopub} and RAPPOR~\cite{erlingsson2014rappor}. To handle categorical data with arbitrary number of possible values, the $k$-RR~\cite{kairouz2014extremal} is proposed. In typical RR, each user shares his answer truthfully with probability $p$ and provide the opposite answer with $1-p$. To achieve $\epsilon$-LDP, the probability $p$ is set by:
	\begin{align}
	p = \frac{\exp(\epsilon)}{\exp(\epsilon)+1}.
	\end{align}
	
	Let $f_r$ denote the proportion of positive (resp. negative) answers received by the aggregator, the frequency of positive (resp. negative) answers before perturbing can be estimated by:
	\begin{align}
	\label{eq: rr adjust}
	f^* = \frac{p-1+f_r}{2p-1},
	\end{align}
	then $f^*$ is an unbiased estimator of $f$. 

	Recently, the numerical value perturbation under LDP for mean estimation has been addressed in the literature. We briefly introduce the Harmony~\cite{nguyen2016collecting} and Piecewise mechanism~\cite{wang2019collecting}.
	
	\subsubsection{Harmony.}
	Nguy{\^e}n \textit{et al.}~\cite{nguyen2016collecting} proposed \textbf{Harmony} for collecting and analyzing data from smart device users. Shown as Alg.~\ref{alg:duchi's mechanism},  Harmony contains three steps: discretization, perturbation and adjusting. The discretization is used to generate a discretized value in \{-1, 1\}, then Randomized Response is applied to achieve $\epsilon$-LDP. At last, to output an unbiased value, the perturbed value is adjusted. 
	
	\begin{algorithm}[!h] 
		\caption{Harmony~\cite{nguyen2016collecting} for Mean Estimation.} 
		\label{alg:duchi's mechanism}
		\begin{algorithmic}[1] 
			\REQUIRE value $v \in [-1, 1]$ and privacy budget $\epsilon$.
			\ENSURE discretized value $x^* \in \{-\frac{e^\epsilon+1}{e^\epsilon-1}, \frac{e^\epsilon+1}{e^\epsilon-1}\}$
			\STATE Discretize value to $v^*\in\{-1, 1\}$ by:
			\begin{align}
			v^* =\operatorname{Dis}(v) = \begin{cases}
			-1 \quad &\text{with probability } \frac{1-v}{2}\\
			1 \quad &\text{with probability } \frac{1+v}{2}
			\end{cases}\nonumber
			\end{align}
			\STATE Perturb $v^*$ by using randomized response:
			\begin{align}
			v^* = \begin{cases}
			v^* \quad &\text{with probability } \frac{e^\epsilon}{e^\epsilon+1}\\
			-v^* \quad &\text{with probability } \frac{1}{e^\epsilon+1}
			\end{cases}\nonumber
			\end{align}
			\STATE Adjusted the perturbed by:
			\begin{align}
			v^* = \frac{e^\epsilon+1}{e^\epsilon-1} \cdot v^*\nonumber
			\end{align}
			\RETURN $v^*$
		\end{algorithmic}
	\end{algorithm}
	
	\subsubsection{Piecewise Mechanism.}
	The Piecewise Mechanism (PM)~\cite{wang2019collecting} is another perturbation solution for mean estimation. Unlike the Harmony, the output domain of PM is continuous from $-\frac{\exp (\epsilon / 2)+1}{\exp (\epsilon / 2)-1}$ to $\frac{\exp (\epsilon / 2)+1}{\exp (\epsilon / 2)-1}$. The PM is used for collecting a single numeric attribute under LDP. Based on PM,~\cite{wang2019collecting} also build a Hybrid Mechanism (HM) for mean estimation. The PM and HM obtain higher result accuracy compared to existing methods.

\subsection{Problem Definition}
	
	This paper researches the problem of data collecting and analyzing while considering users' privacy preferences in the context of LDP. For simplicity, we assume that each user holds one single value $v_i$.
	
	\begin{table}[!h]
		\centering 
		\caption{Notations.} 
		\begin{tabular}{|c|c|}
			\hline
			\textbf{Symbol} & \textbf{Description}\\
			\hline
			$\mathcal{U} = \{u_1, u_2, ..., u_n\}$ & the set of users, where $n=|\mathcal{U}|$\\
			\hline 
			$v_i$ &  value of user $u_i$, $v_i \in [-1,1]\cup \{\bot\}$\\
			\hline
			$\epsilon_u^i$ & privacy demand of $u_i$\\
			\hline
			$\epsilon$ & privacy budget of perturbation mechanism\\
			\hline
			$p$ & $p = e^{\epsilon} / (e^{\epsilon}+1)$\\
			\hline
		\end{tabular}
	\end{table}
	
	\textbf{Modeling users' privacy preferences.} As detailed in the introduction part, for one single investigating question, different users have different privacy preferences. Without loss of generality, we use $\epsilon_u^i$ to describe the privacy-preserving preferences of $u_i$ and we assume that user $u_i$ only collaborates with the data collector when the provided privacy-preserving level is higher than expected (which is $\epsilon \le \epsilon_u^i$). When the provided privacy-preserving level is not as expected ($\epsilon > \epsilon_u^i$), the user $u_i$ provides a null-value (represented by $v_i = \bot$) instead of the fake answer for perturbation.
	
	After perturbation, data from users' side are collected by an untrusted aggregator, who wants to learn some statistical information from all users, especially the rate of null-value and the mean value of all users.
	
	\begin{definition}[Mean of missing data]
		\label{def: mean}
		For a list of values $\boldsymbol{v} = \{v_1, v_2, ..., v_n\}$ where each value $v_{i:i\in [n]}$ from user $u_i$ is in domain $ [-1,1] \cup \{\bot\}$, the missing rate and the mean of $\textbf{v}$ is defined as:
		\begin{align}
		mr = \frac{\#\{v_i | v_i = \bot\}}{n}, \quad m = \frac{\sum_{v_i\not=\bot} v_i}{\# \{v_i | v_{i} \not= \bot\}}.
		\end{align}
	\end{definition}

	Also, when $\forall i: \epsilon_u^i \ge \epsilon$, the estimation of mean of missing data turns to be the tradition mean estimation problems:
	
	\begin{definition}[Mean of values]
		For a list of values $\boldsymbol{v} = \{v_1, v_2, ..., v_n\}$ where each value $v_{i\in[n]}$ is in domain $[-1, 1]$. The mean of $\textbf{v}$ is defined as:
		\begin{align}
		m = \frac{\sum_{i\in[n]} v_i}{n}.
		\end{align}
	\end{definition}
	
\section{BiSample: Bidirectional Sampling Technique}
\label{sec: bisample}

	Before presenting solution for missing data perturbation, we first propose a bidirectional sampling technique, referred to as the \textit{BiSample Mechanism}. The BiSample mechanism takes a value $v \in [-1, 1]$ as input and outputs a perturbed tuple $\langle s, b \rangle$ where $s$ represents the sampling direction and $b$ represents the sampling result of $v$. Specifically, the BiSample mechanism contains two basic sampling directions, which is defined as:
	
	\begin{itemize}
		\item \textbf{Negative Sampling with LDP}. The negative sampling is used to estimate the frequency of -1 after discretization. The perturbing procedure of negative sampling is: 
		\begin{align}
		\Pr[b=1] = (2p-1)\cdot \Pr[\operatorname{Dis}(v)=-1] + (1-p).
		\end{align}
		\item \textbf{Positive Sampling with LDP}. Like negative sampling, the positive sampling is used to estimate the frequency of 1 after discretization. Notably, the typical RR is positive sampling.
		\begin{align}
		\Pr[b=1] = (2p-1)\cdot \Pr[\operatorname{Dis}(v)=1] + (1-p).
		\end{align}
	\end{itemize}
	
	\begin{algorithm}[t] 
		\caption{\textit{BiSample}$(v, \epsilon)$: Bidirectional Sampling Mechanism} 
		\label{alg: bisample}
		\begin{algorithmic}[1] 
			\REQUIRE a value $v\in[-1,1]$, privacy budget $\epsilon$.
			\STATE sample a uniformly variable $s\in\{0,1\}$ representing the sampling direction.
			\IF {$s=0$}
			\STATE use \textbf{Negative Sampling}: generate a Bernoulli variable $b$ with:
			\begin{align}
			\Pr[b=1] = \frac{1-\exp({\epsilon})}{1+\exp({\epsilon})}\cdot \frac{v}{2} + \frac{1}{2}. \nonumber
			\end{align}
			\ELSE
			\STATE use \textbf{Positive Sampling}: generate a Bernoulli variable $b$ with:
			\begin{align}
			\Pr[b=1] = \frac{\exp({\epsilon})-1}{\exp({\epsilon})+1}\cdot \frac{v}{2} + \frac{1}{2}. \nonumber
			\end{align}
			\ENDIF
			
			\RETURN $s, b$.
		\end{algorithmic}
	\end{algorithm}
	
	Assuming the input domain is $[-1, 1]$, Algorithm~\ref{alg: bisample} shows the pseudo-code of BiSample. Without loss of generality, when the input domain is $[L, U]$, the user (i) computes $v' = \frac{2}{U-L}\cdot v + \frac{L+U}{L-U}$, (ii) perturbs $v'$ using the BiSample mechanism, and (iii) shares $\langle s, \left(\frac{U-L}{2}\right)\cdot b +\frac{U+L}{2}\rangle$ with the aggregator, where $s$ denotes the sampling method and $b$ is the sampling result of $v'$. In Algorithm~\ref{alg: bisample}, Lines 2-3 show the negative sampling process and Lines 5-6 denote the positive sampling. We prove that the combination of positive and negative sampling satisfies $\epsilon$-LDP.
	
	\begin{theorem}
		\label{thm: bisample}
		The BiSample mechanism $\mathcal{M} = \operatorname{BiSample}(\cdot)$ guarantees $\epsilon$-LDP.
		\begin{proof}
			For any $t_1, t_2 \in [-1,1]$ and output $o \in \operatorname{Range}(\mathcal{M})$, we have:
			\begin{align}
			&
			\ln\max_{t_1, t_2\in[-1,1], o\in\operatorname{Range}(\mathcal{M})} \frac{\Pr[\mathcal{M}(t_1)=o]}{\Pr[\mathcal{M}(t_2)=o]}\nonumber\\
			&= \ln \max_{t_1, t_2\in[-1,1] , b\in \{0,1\}} \frac{\Pr[\mathcal{M}(t_1)=\langle 0, b\rangle ]}{\Pr[\mathcal{M}(t_2)=\langle 0, b\rangle]}\nonumber\\
			&=\ln \frac{\max_{t_1 \in [-1,1]} \Pr[\mathcal{M}(t_1)=\langle 0, 0 \rangle]}{\min_{t_2 \in [-1,1]} \Pr[\mathcal{M}(t_2)=\langle 0, 0 \rangle]}\nonumber\\
			&= \ln \left(\frac{\exp(\epsilon)}{2(\exp(\epsilon)+1)}/\frac{1}{2(\exp(\epsilon)+1)}\right) = \epsilon.
			\end{align}
			
			According to the definition of LDP, the BiSample achieves $\epsilon$-LDP.
		\end{proof}
	\end{theorem}
	
	With the BiSample perturbation, a value $v_i$ in the input domain is perturbed into a two-bit tuple $\mathcal{M}_{\operatorname{BiSample}}(v_i) = \langle s_i, b_i \rangle$. The result is two-fold. First, the $s_i$ indicates whether the sampling mechanism is positive sampling or not. Second, the $b_i$ represents the sampling value with correspond sampling mechanism. For the aggregator, let $\mathcal{R}=\{\langle s_1, b_1\rangle, \langle s_2, b_2\rangle, ... \langle s_n, b_n\rangle\}$ be the perturbed data received from all the users and $f_{\operatorname{POS}}$ (resp. $f_{\operatorname{NEG}}$) be the aggregated frequency of positive sampling (resp. negative sampling), which is given by:
	\begin{align}
	\label{equ:fpos}
	f_{\operatorname{POS}} &= \frac{\#\{\langle s_i, b_i\rangle| \langle s_i, b_i\rangle = \langle 1,1\rangle, \langle s_i, b_i \rangle \in \mathcal{R}\}}{\#\{\langle s_i, b_i\rangle | s_i=1, \langle s_i, b_i\rangle \in \mathcal{R}\}},\\
	\label{equ:fneg}
	f_{\operatorname{NEG}} &= \frac{\#\{\langle s_i, b_i\rangle| \langle s_i, b_i\rangle = \langle 0,1\rangle, \langle s_i, b_i \rangle \in \mathcal{R}\}}{\#\{\langle s_i, b_i\rangle | s_i=0, \langle s_i, b_i\rangle \in \mathcal{R}\}}.
	\end{align}
	
	\begin{theorem}
	\label{thm: unbiased of basic BiSample}
		$m^*$ is an unbiased estimator of $m=\frac{1}{n}\sum_{i:i\in[n]}v_i$, where $m^*$ is given by:
		\begin{align}
		m^* = \frac{1}{2p-1}\left(f_{\operatorname{POS}} - f_{\operatorname{NEG}}\right).
		\end{align}
		\begin{proof}
			Firstly, the $m^*$ can be represented by:
			\begin{align}
			\mathbb{E}[m^*] &= \mathbb{E}\left[\frac{1}{2p-1} \left(f_{\operatorname{POS}} - f_{\operatorname{NEG}}\right)\right]\nonumber\\
			&= \mathbb{E}\left[\frac{1}{2p-1}\left((f_{\operatorname{POS}} + p-1) - (f_{\operatorname{NEG}}+p-1)\right)\right] \nonumber\\
			&= \left(\mathbb{E}\left[ \frac{f_{\operatorname{POS}} + p-1}{2p-1}\right] - \mathbb{E}\left[\frac{f_{\operatorname{NEG}} + p-1}{2p-1}\right]\right).
			\end{align}
			
			Then, according to Eq.~(\ref{eq: rr adjust}), the $ \frac{f_{\operatorname{POS}} + p-1}{2p-1}$ (resp. $\frac{f_{\operatorname{NEG}} + p-1}{2p-1}$) represents the estimated frequency of number $1$ (resp. $-1$) before perturbation. According to the bidirectional sampling process, we then have:
			\begin{align}
			\mathbb{E}[m^*] &= \frac{1}{n} \cdot \mathbb{E} \left[\#\{i|\operatorname{Dis}(v_i)=1\}\right] - \frac{1}{n} \cdot  \mathbb{E}\left[\#\{i| \operatorname{Dis}(v_i)=-1\}\right] \nonumber\\
			&= \frac{1}{n} \cdot \sum_{i\in [n]} v_i = m.
			\end{align}
			
			We then conclude that $m^*$ is unbiased. Also, the variance of BiSample is given by:
			\begin{align}
			\operatorname{Var}[m^*] &= \mathbb{E}[(m^*)^2] - (\mathbb{E}[m^*])^2\nonumber\\
			&= \left(\frac{\exp(\epsilon)+1}{\exp(\epsilon)-1}\right)^2 - m^2.
			\end{align}
		\end{proof}
	\end{theorem}
	
	Therefore, the worst-case variance of the BiSample mechanism equals $\left(\frac{e^\epsilon+1}{e^\epsilon-1}\right)^2$, which is the same as the Harmony solution. Normally, when using perturbation in $d$-dimensional data with $\epsilon$-LDP guarantee, the maximum difference between the true mean and the estimated mean is bounded with high probability. Shown as Theorem~\ref{thm: error bound}, the proof is similar to the one in~\cite{nguyen2016collecting}.


	\begin{theorem}
	\label{thm: error bound}
		For $j \in [d]$, let $m_j^*$ denote the estimator of $m_j = \frac{1}{n} \sum_{i\in[n]} v_{i,j}$ by the BiSample mechanism. With at least probability $1-\beta$, we have:
		\begin{align}
		\left|m_j^* - m_j\right| = O\left(\frac{\sqrt{d\cdot \operatorname{log}(d/\beta)}}{\sqrt{n}\cdot \epsilon}\right).
		\end{align}
	\end{theorem}
	
\section{Using BiSample for Missing Data Perturbation}
\label{sec: bisample-md}

	\begin{figure}[htbp]
		\centering
		\includegraphics[width=120mm]{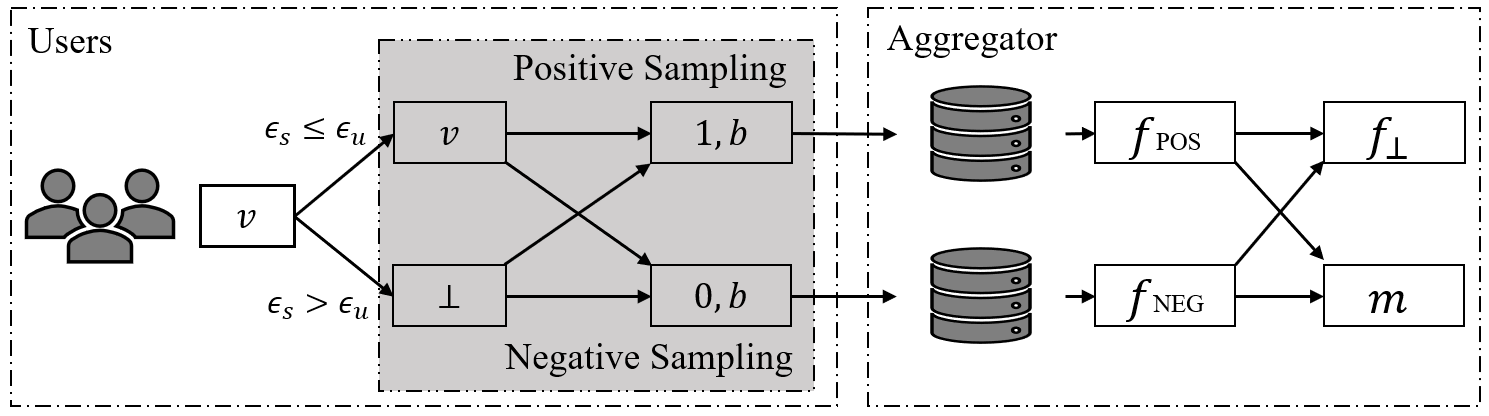}
		\caption{The BiSample-MD Framework.}
		\label{fig: bisample-md}
	\end{figure}
	
	The proposed BiSample mechanism uses a bi-directional sampling technique for numerical value perturbation. However, it cannot handle the \textbf{fake answer} situation. In this section, we consider a providing null-value procedure and propose the BiSample-MD framework that extends the BiSample for missing data.
	
	Fig.~\ref{fig: bisample-md} illustrates the BiSample-MD model. We use $\epsilon_u^i$ to represent the privacy preference of $u_i$ and use $\epsilon$ to represent the privacy budget of the perturbation mechanism provided by the aggregator. Before perturbing value locally, each user uses the $PV(v, \epsilon_u, \epsilon)$ (shown as Algorithm~\ref{alg: prepare value}) to decide whether using the real value or not. When the privacy-preserving level of perturbation mechanism is higher than user's expectation, the $PV(v_i,\epsilon_u^i, \epsilon)$ returns the real value $v'=v$. Otherwise, the $PV(\cdot)$ returns a null-value $v'=\bot$.

	\begin{algorithm}[!h] 
		\caption{\textit{PV}($v, \epsilon_u, \epsilon$): Prepare Value.}
		\label{alg: prepare value}
		\begin{algorithmic}[1] 
			\REQUIRE user's value $v \in [-1,1]$ and user's expected privacy budget $\epsilon_u$, system privacy budget $\epsilon$.
			\IF {$\epsilon \le \epsilon_u$}
			\RETURN $v$.
			\ELSE
			\RETURN $\bot$.
			\ENDIF
		\end{algorithmic}
	\end{algorithm}
	
	Then $v'$ is used for perturbing instead of $v$, the domain of $v'$ is $[-1,1]\cup \{\bot\}$. We then design the BiSample-MD algorithm for perturbing $v'$. Even though the input domain is different from that of BiSample, we still design the output domain to be $s\in\{0,1\}, b \in\{0,1\}$. The BiSample-MD perturbation process is detailed in Algorithm~\ref{alg: bisample-md}. Like BiSample, the BiSample-MD also contains positive sampling and negative sampling. When $v'=\bot$, both the positive and negative sampling all sample $b=1$ with probability $1/(\exp(\epsilon)+1)$. The following theorem shows that the BiSample-MD algorithm satisfies $\epsilon$-LDP.

	\begin{algorithm}[!h] 
		\caption{\textit{BiSample-MD$(v, \epsilon_u, \epsilon)$}: BiSample for Missing Data.}
		\label{alg: bisample-md}
		\begin{algorithmic}[1] 
			\REQUIRE user's value $v \in [-1,1]$ and user's expected privacy budget $\epsilon_u$, system privacy budget $\epsilon$.
			\STATE $v'=PV(v, \epsilon_u, \epsilon)$ 
			\STATE sample a uniformly variable $s\in\{0,1\}$ representing the sampling direction.
			\IF {$s=0$}
			\STATE Generate a Bernoulli variable $b$ with:
			\begin{align}
				\Pr[b=1] = \begin{cases}
				\frac{1-\exp(\epsilon)}{1+\exp(\epsilon)}\cdot \frac{v'}{2} + \frac{1}{2} &\quad \text{if } v'\in[-1,1];\\
				\frac{1}{\exp(\epsilon)+1} &\quad \text{if } v'=\bot.
				\end{cases}\nonumber
			\end{align}
			\ENDIF
			\IF {$s=1$}
			\STATE Generate a Bernoulli variable $b$ with:
			\begin{align}
				\Pr[b=1] = \begin{cases}
				\frac{\exp(\epsilon)-1}{\exp(\epsilon)+1}\cdot \frac{v'}{2} + \frac{1}{2} &\quad \text{if } v'\in[-1,1];\\
				\frac{1}{\exp(\epsilon)+1} &\quad \text{if } v'=\bot.
				\end{cases}\nonumber
			\end{align}
			\ENDIF
			\RETURN $s, b$
		\end{algorithmic}
	\end{algorithm}

	\begin{theorem}
		\label{thm: bisample-md}
		Alg.~\ref{alg: bisample-md} achieves $\epsilon$-LDP.	
		\begin{proof}
			Shown in Theorem~\ref{thm: bisample}, it is proven that when $\epsilon_u^i > \epsilon$, the BiSample-MD mechanism is $\epsilon$-LDP. In this way, we only need to consider the situation when the null-value occurs. Without loss of generality, we assume $t_1 = \bot$. we have:
			\begin{align}
			&
		 \max_{t_2\in[-1,1], s\in \{0,1\},b \in \{0,1\}} \left\{\frac{\Pr\left[\mathcal{M}(\bot)=\langle s, b\rangle\right]}{\Pr\left[\mathcal{M}(t_2)=\langle s, b\rangle\right]}, \frac{\Pr\left[\mathcal{M}(t_2)=\langle s, b\rangle\right]}{\Pr\left[\mathcal{M}(\bot)=\langle s, b\rangle\right]} \right\}\nonumber\\
			&=\max_{t_2\in [-1,1], b\in\{0,1\}} \frac{\Pr\left[\mathcal{M}(\bot)=\langle 0, b\rangle\right]}{\Pr\left[\mathcal{M}(t_2)=\langle 0, b\rangle\right]} = \frac{\Pr[\mathcal{M}(\bot) = \langle 0,0 \rangle]}{\min_{t_2 \in [-1,1]} \Pr[\mathcal{M}(t_2)=\langle 0, 0 \rangle]}.\nonumber
			\end{align}

			
			According the perturbation mechanism, the numerator is given by:
			\begin{align}
			\label{equ: budget_1}
			\Pr\left[\mathcal{M}(\bot)=\langle 0, 0\rangle\right]=\Pr[s=0]\cdot \Pr[b=0] = \frac{1}{2}\cdot \frac{\exp(\epsilon)}{1+\exp(\epsilon)},
			\end{align}
			while the denominator can be calculated by:
			\begin{align}
			\label{equ: budget_2}
			\min_{t_2 \in [-1,1]} \Pr[\mathcal{M}(t_2)=\langle 0, 0 \rangle] = \Pr[\mathcal{M}(-1)=\langle 0, 0 \rangle] = \frac{1}{2}\cdot \frac{1}{\exp(\epsilon)+1}.
			\end{align}
			
			According to Eq.~(\ref{equ: budget_1}) and Eq.~(\ref{equ: budget_2}), the privacy budget is bounded by $\epsilon$ when the value is a null-value. To sum up, the BiSample-MD algorithm is $\epsilon$-LDP.
		\end{proof}
	\end{theorem}

	The perturbed data are then collected by the aggregator. Let $s$ be the sum of values provided truthfully, which is given by $s = \sum_{i: \epsilon_u^i < \epsilon} v_i$ and $f_\bot$ be the fraction of users who provide a null-value for perturbation, which is given by $f_\bot= \#\{i: \epsilon_u^i < \epsilon\} / n$. Then we can estimate $s$ and $f_\bot$ by:
	\begin{align}
	s^* = \frac{n}{2p-1}\cdot (f_{\operatorname{POS}}-f_{\operatorname{NEG}}),\indent
	f_\bot^* = \frac{1 - f_{\operatorname{POS}} - f_{\operatorname{NEG}}}{2p-1},
	\end{align}
	where the $f_{\operatorname{POS}}$ and $f_{\operatorname{NEG}}$ are defined in Equations~\ref{equ:fpos} and~\ref{equ:fneg}. The correctness of the estimation is given by the following theorem.
	\begin{theorem}
		\label{theorem: unbiased of bisample-md}
		$s^*$ and $f_{\bot}^*$ are unbiased estimators of $s$ and $f_\bot$.
		\begin{proof}
			The main intuition behind this theorem is that in the positive sampling process, the perturbed result only contains whether the value is 1 or not. Under such principle, there is no difference between $v=-1$ of $v=\bot$ when using positive sampling. Thus, following the proof of Theorem.~\ref{thm: unbiased of basic BiSample}, it is easy to prove that $s^*$ is unbiased. For $f_\bot$ we have:
			\begin{align}
			\mathbb{E}[f_{\bot}^*] &= \mathbb{E}[\frac{1 - f_{\operatorname{POS}} - f_{\operatorname{NEG}}}{2p-1}]\noindent\nonumber\\
			&= 1 - \mathbb{E}[\frac{f_{\operatorname{POS}} + p-1}{2p-1}] - \mathbb{E}[\frac{f_{\operatorname{NEG}} + p-1}{2p-1}]\noindent\nonumber\\
			&= f_\bot.		
			\end{align}
			
			Thus, both $s^*$ and $f_\bot^*$ are unbiased.
		\end{proof}
	\end{theorem}
	
	With the unbiased estimator of the sum of values $\sum_{v_{i}\not=\bot} v_{i}$ and the unbiased estimator of the missing rate, we can then estimate the mean by:
	\begin{align}
	m^* = \frac{s^*}{n\cdot (1 - f_\bot^*)}
	= \frac{ f_{\operatorname{POS}} - f_{\operatorname{NEG}} }{f_{\operatorname{POS}} + f_{\operatorname{NEG}} + 2p-2}.
	\end{align}
		
\section{Experiments}
\label{sec: experiment}

	In the experimental part, we empirically evaluate the proposed mechanisms. 
	
	\subsection{Experimental Settings}
	\textbf{Datasets}. To evaluate the proposed mechanisms, we first generated three synthetic datasets: the GAUSS follows Gaussian distribution with location $\mu=0.5$ and scale $\sigma=0.1$, the EXP dataset follows Exponential distribution with scale $0.1$ and the UNIFORM dataset follows uniform distribution. Each dataset contains $10^5$ users. We also use the ADULT dataset~\cite{adultdataset} for evaluation. We extract the Age attribution and regularize each value to $[-1,1]$ for mean estimation.

	\begin{table}[!h]
		\centering 
		\caption{Dataset Description.} 
		\begin{tabular}{|m{2.5cm}<{\centering}|m{2.5cm}<{\centering}|m{2.5cm}<{\centering}|m{2.5cm}<{\centering}|}
			\hline
			\textbf{Dataset} & \textbf{Distribution} & \textbf{$\#$ Instances} & \textbf{Mean Value}\\
			\hline
			EXP & exponential  &$10^5$ & -0.831	\\
			\hline
			GAUSS & Gaussian & $10^5$ & 0.499	\\
			\hline
			UNIFORM & uniform & $10^5$ &-0.001	\\
			\hline
			ADULT & - & $32561$ & -0.409	\\
			\hline
		\end{tabular}
	\end{table}
	
	\vspace{5pt}
	\noindent\textbf{Methodology for null-value perturbation}. In terms of mean estimation, we compare BiSample with the Harmony~\cite{nguyen2016collecting} and the Piecewise mechanism (PM~\cite{wang2019collecting}). For missing data perturbation, we encode the original data to a key-value format. The real value $v$ is represented by $\langle 1,v \rangle$ and the null-value $v=\bot$ is represented by $\langle 0, -\rangle$. We then use PrivKVM~\cite{qingqing2019privkv} for the missing rate estimation and mean estimation. The missing rate is given by $f_\bot = 1 - f_k$, where $f_k$ is the frequency of key given by PrivKVM. We use one real iteration and five virtual iterations.
	
	\vspace{5pt}
	\noindent\textbf{Utility metric}. All experiments are performed 100 times repeatedly. We evaluate the performance of missing rate ($mr$) estimation and mean estimation ($m$) by the average absolute error and variance, which are defined by ($T=100$):
	\begin{align}
	\begin{cases}
		\operatorname{AE}(mr) = \frac{1}{T} \sum |f_\bot - f_\bot^*|, &\operatorname{AE}(m) = \frac{1}{T} \sum |m - m^*|;\\
		\operatorname{Var}(mr) = \frac{1}{T} \sum (f_\bot - f_\bot^*)^2, &\operatorname{Var}(m) = \frac{1}{T} \sum (m - m^*)^2.
	\end{cases}
	\end{align}
	where $f_\bot$ and $m$ (resp. $f_\bot^*$ and $m^*$) are the true (resp. estimated) missing rate and the mean value.


	\begin{figure}[t]
		\centering
		\footnotesize
		\begin{tabular}{cc}							    	
			\includegraphics[width=0.5\textwidth]{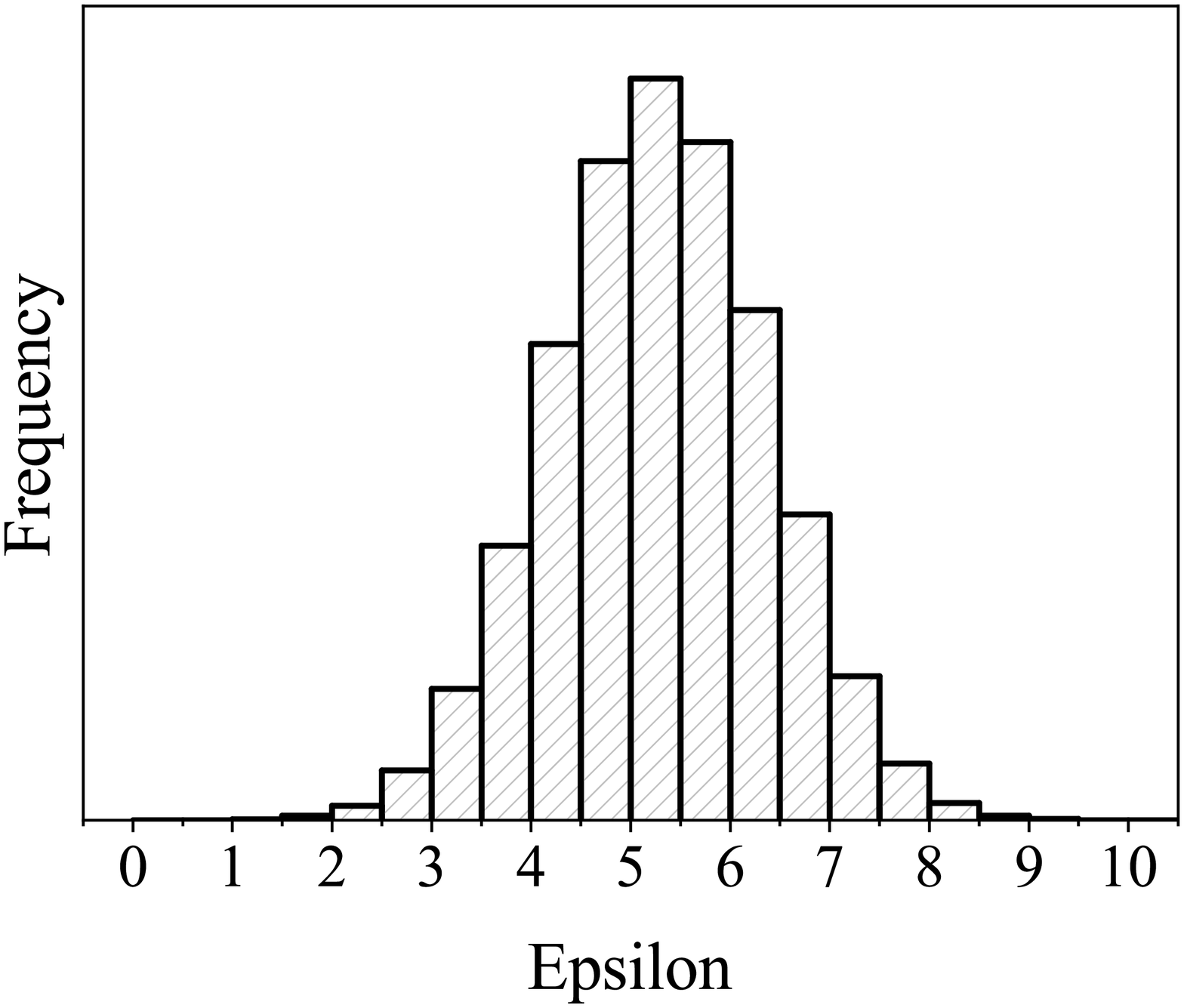} & 
			\includegraphics[width=0.5\textwidth]{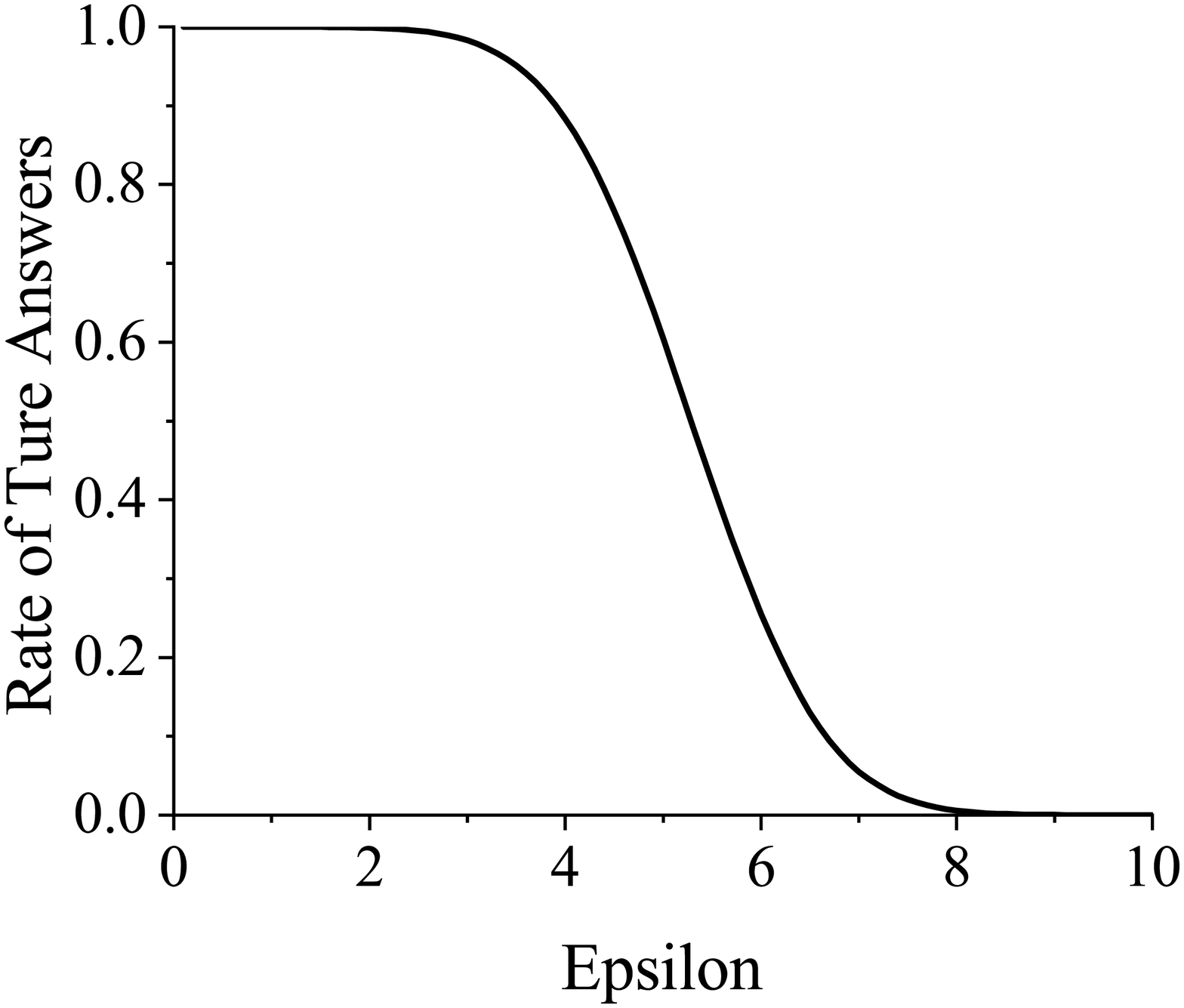}
			\\ 
			(a) Users' Privacy Preferences Distribution. &
			(b) Rate of True Answers varying $\epsilon$. 
		\end{tabular}
		\caption{The Impact of Users' Privacy Preferences to Rate of True Answers.}
		\label{fig: user privacy preferences}
	\end{figure}

\subsection{Varying User Behavior}

	In this part of experiments, we consider the task of collecting an 1-dimensional value from each user while considering users' privacy preferences. Since no existing solution researched the distribution of users' privacy preferences, shown as Fig.~\ref{fig: user privacy preferences}(a), we generated Gaussian data with $\mu=5$ and $\sigma=1.5$ as the users' preferences distribution. Fig.~\ref{fig: user privacy preferences}(b) plots the rate of users who would truthfully provide the real value for perturbation according to $\epsilon$. Basically, when the privacy budget $\epsilon$ is small, the privacy-preserving level provided by the perturbation mechanism is high, so most people would like to share their real value. In contrast, with a high $\epsilon$, few people want to use real value for perturbation as the perturbing process is not privacy-preserving enough. As existing solutions forces an input, we consider two kinds of user behaviors when the privacy-preserving level of LDP is lower than users' expectation: the \textbf{TOP} mode and the \textbf{RND} mode. In the TOP mode, users always use the value 1 instead of the real value for perturbation. In the RND mode, each user randomly generates a value, uses it for perturbation and shares the perturbed result. 
	
	\begin{figure}[t]
		\centering
		\footnotesize
		\begin{tabular}{cc}
			\includegraphics[width=0.5\textwidth]{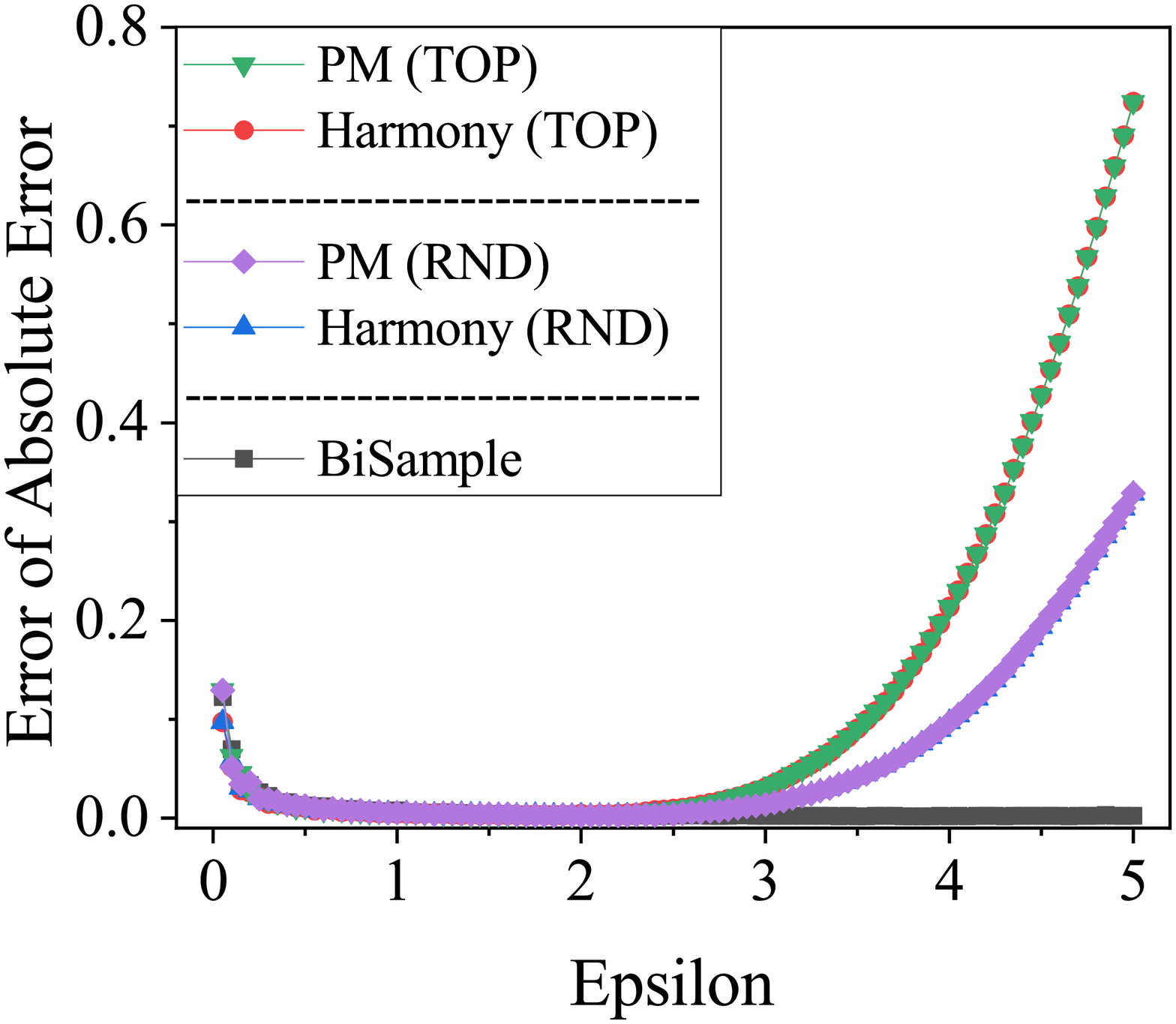} & 
			\includegraphics[width=0.5\textwidth]{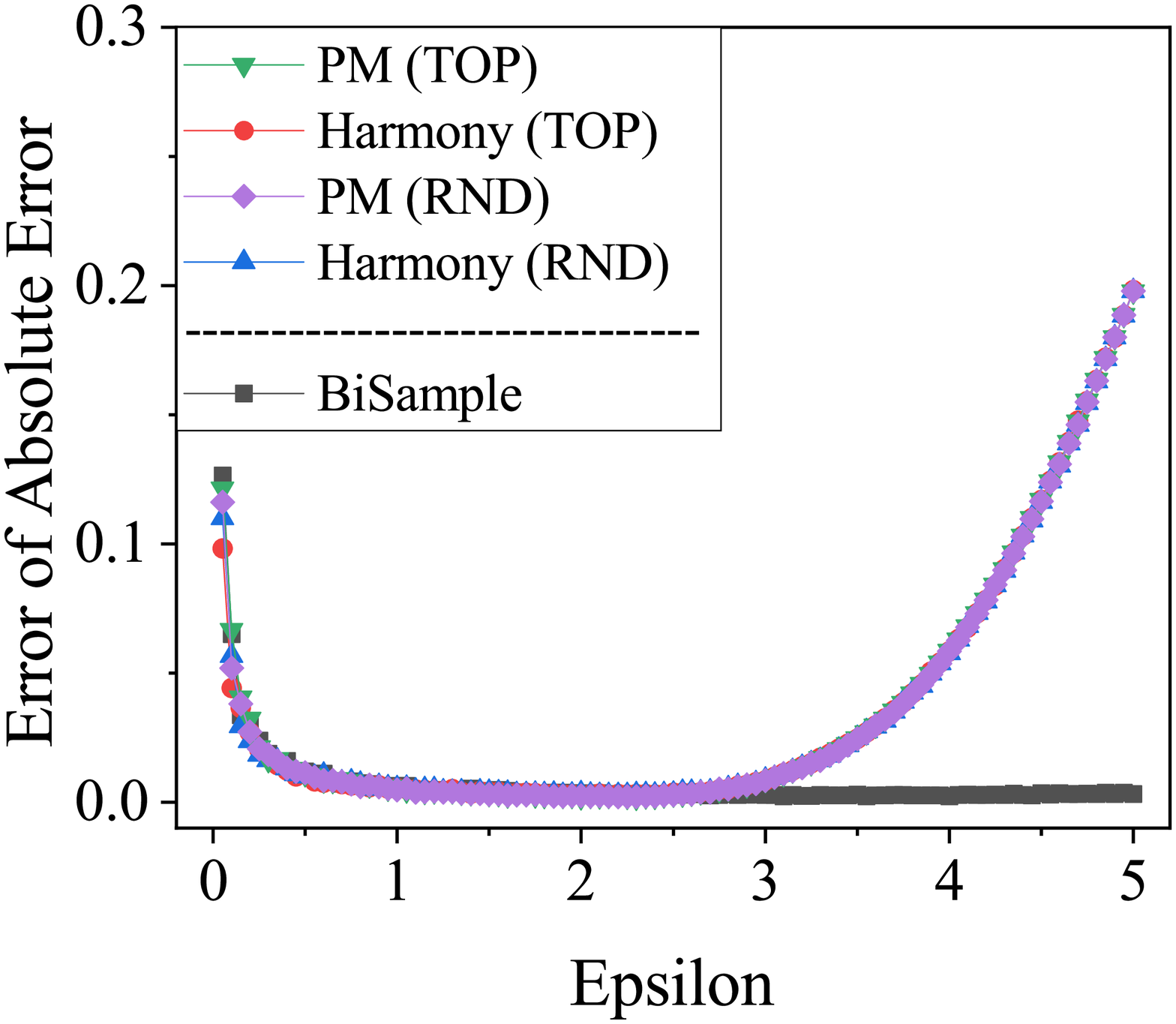}  \\
			(a) EXP Dataset.&
			(b) GAUSS Dataset.
			\\
			\includegraphics[width=0.5\textwidth]{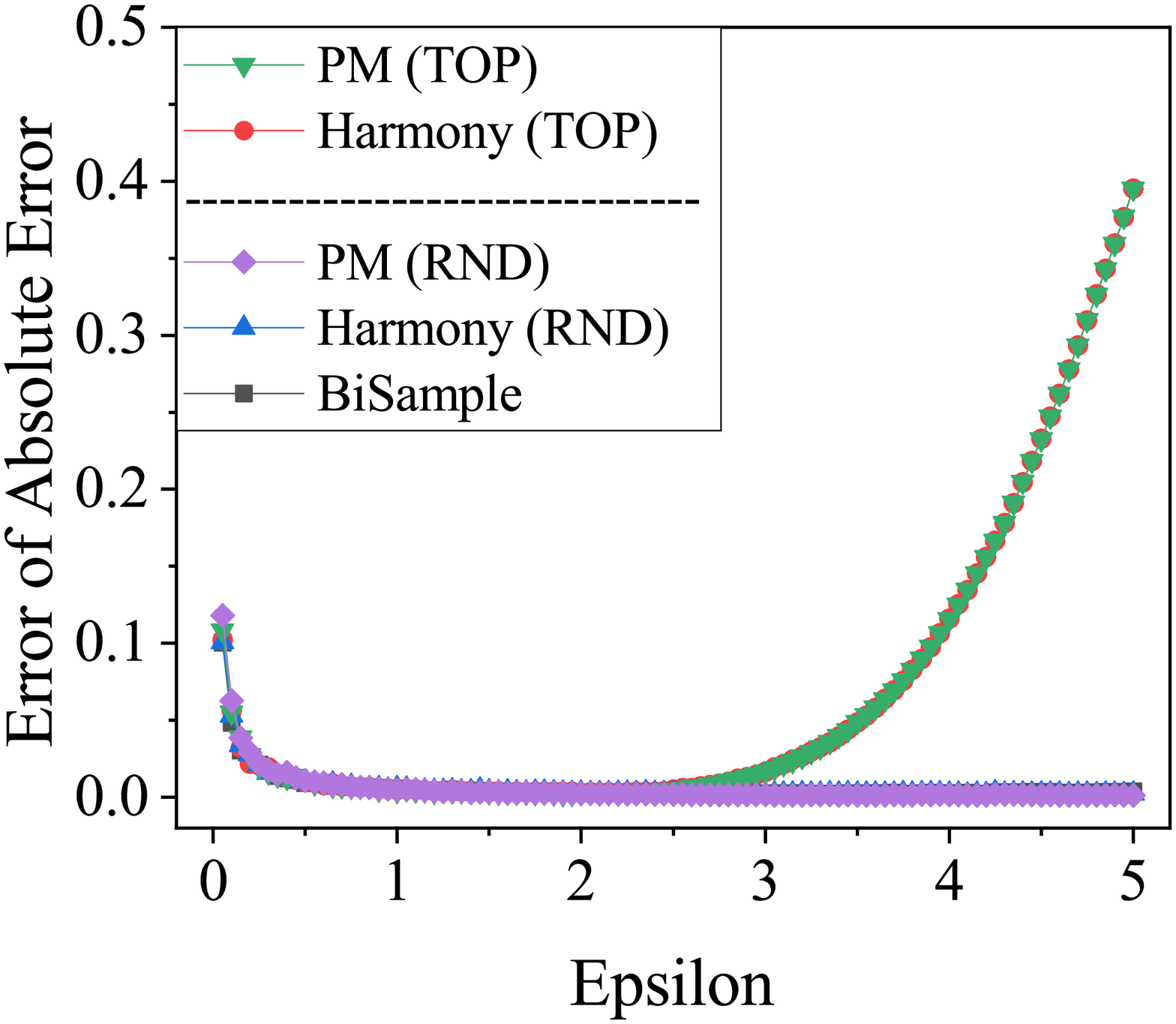} & 
			\includegraphics[width=0.5\textwidth]{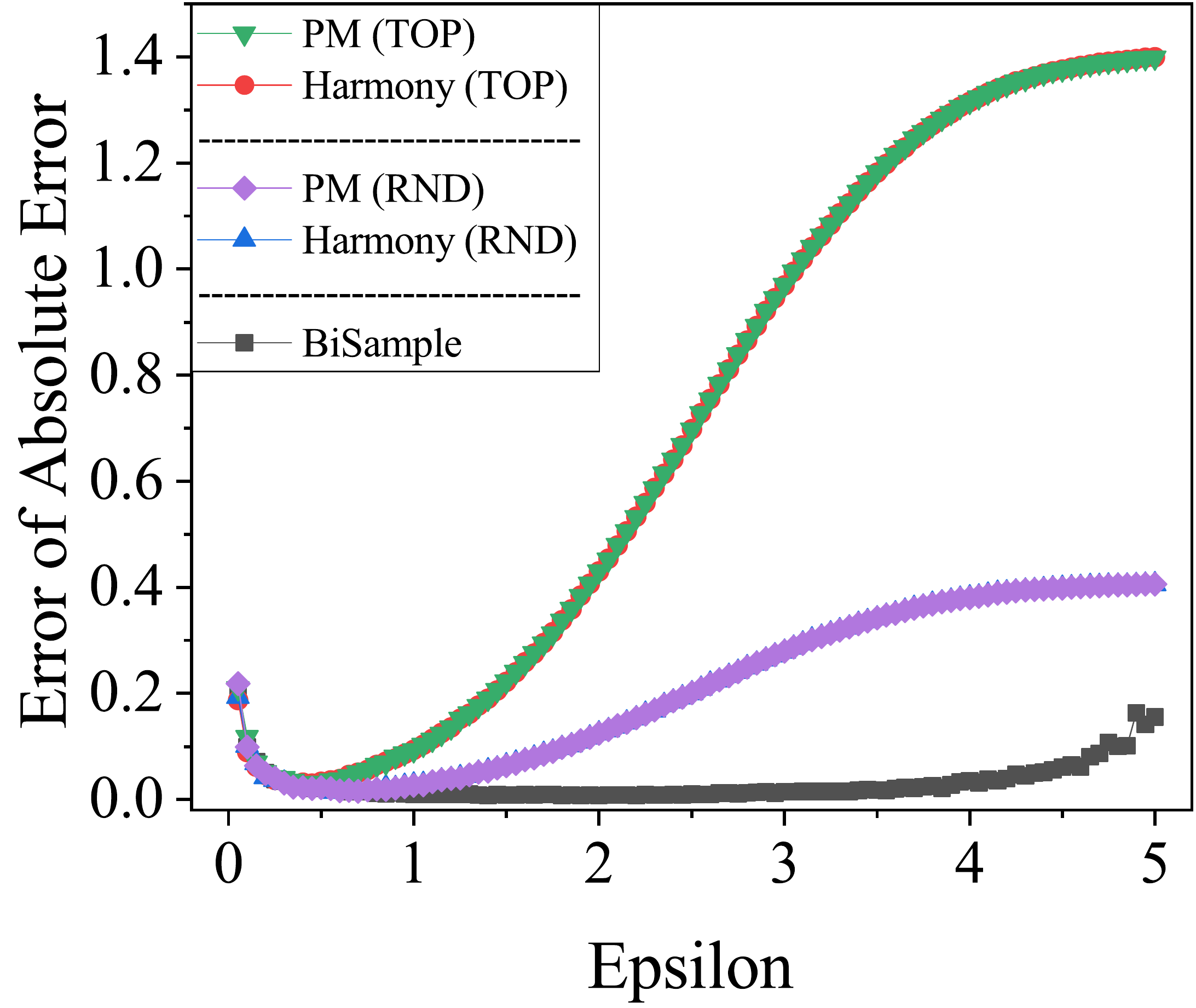}
			\\
			(c) UNIFORM Dataset.&
			(d) ADULT Dataset.
		\end{tabular}
		\caption{Mean Estimation on Different Datasets.}
		\label{fig: cmp_usr}
	\end{figure}
	
	Using the absolute error as utility measurement, Fig.~\ref{fig: cmp_usr} shows the average absolute error over both synthetic datasets and real-world dataset with the change of $\epsilon$, where the TOP and RND represent user behaviors when $\epsilon_u < \epsilon$. For the presentation purpose, methods with similar performance are grouped together in Fig.~\ref{fig: cmp_usr}. We first observe that for the PM and Harmony mechanisms, the performance is close to each other. For these two mechanisms, the influence of user behaviors (TOP or RND) is great. In Fig.~\ref{fig: cmp_usr}(c), it is a coincidence that the performance of RND-based mechanisms are as good as the BiSample mechanism because for uniformly distributed data. This is explainable as randomly generate a fake answer would not affect the mean value statistically.
	
	Usually, in conventional settings without consideration of users' privacy preferences, the error of mean estimation becomes smaller with a larger $\epsilon$, as the privacy-preserving level decreases. However, our simulation shows a different opinion. With the increase of privacy budget, the estimation performance would also become poor. As for users, when the privacy budget is too large, they refuse to provide the real value for perturbation because they think their privacy is not well-guaranteed. Also, in this setting, only the PrivKVM and BiSample mechanisms can estimate the missing rate.

\subsection{Varying Missing Rate}
	\begin{figure}[t]
		\centering
		\footnotesize
		\begin{tabular}{cc}
			\includegraphics[width=0.5\textwidth]{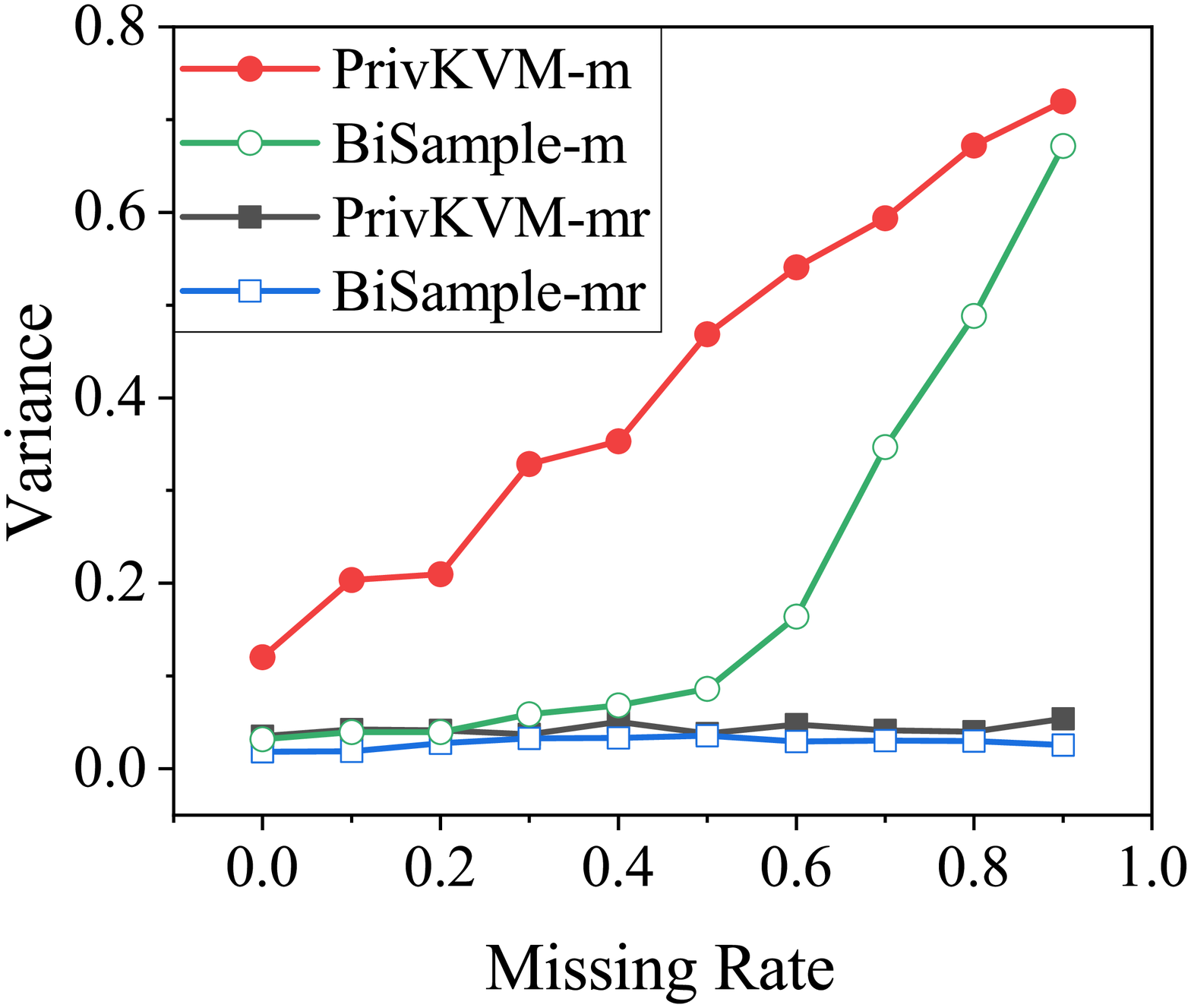} & 
			\includegraphics[width=0.512\textwidth]{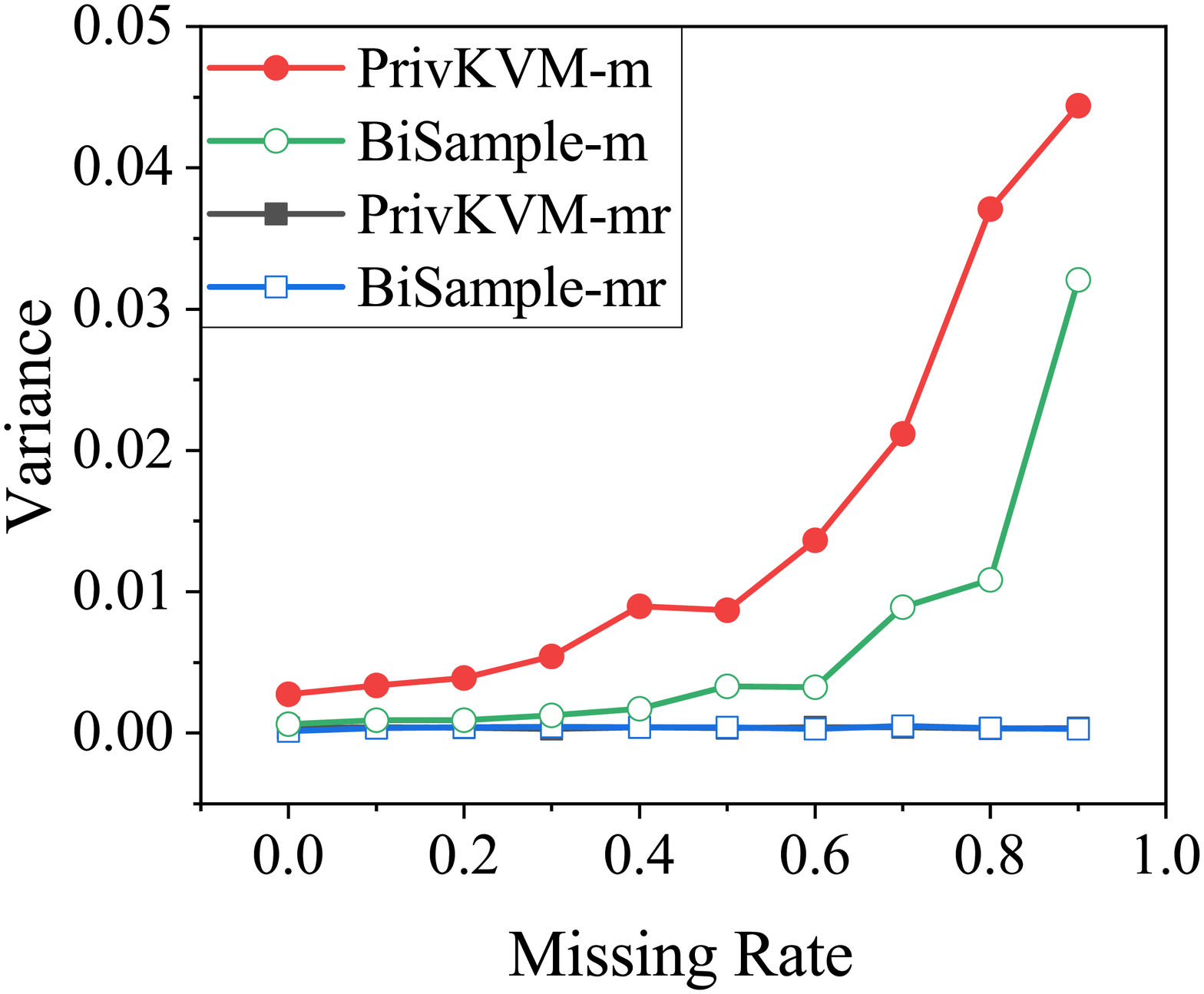}  \\
			(a) EXP ($\epsilon=0.1$). &
			(b) EXP ($\epsilon=1$).
			\\
			\includegraphics[width=0.5\textwidth]{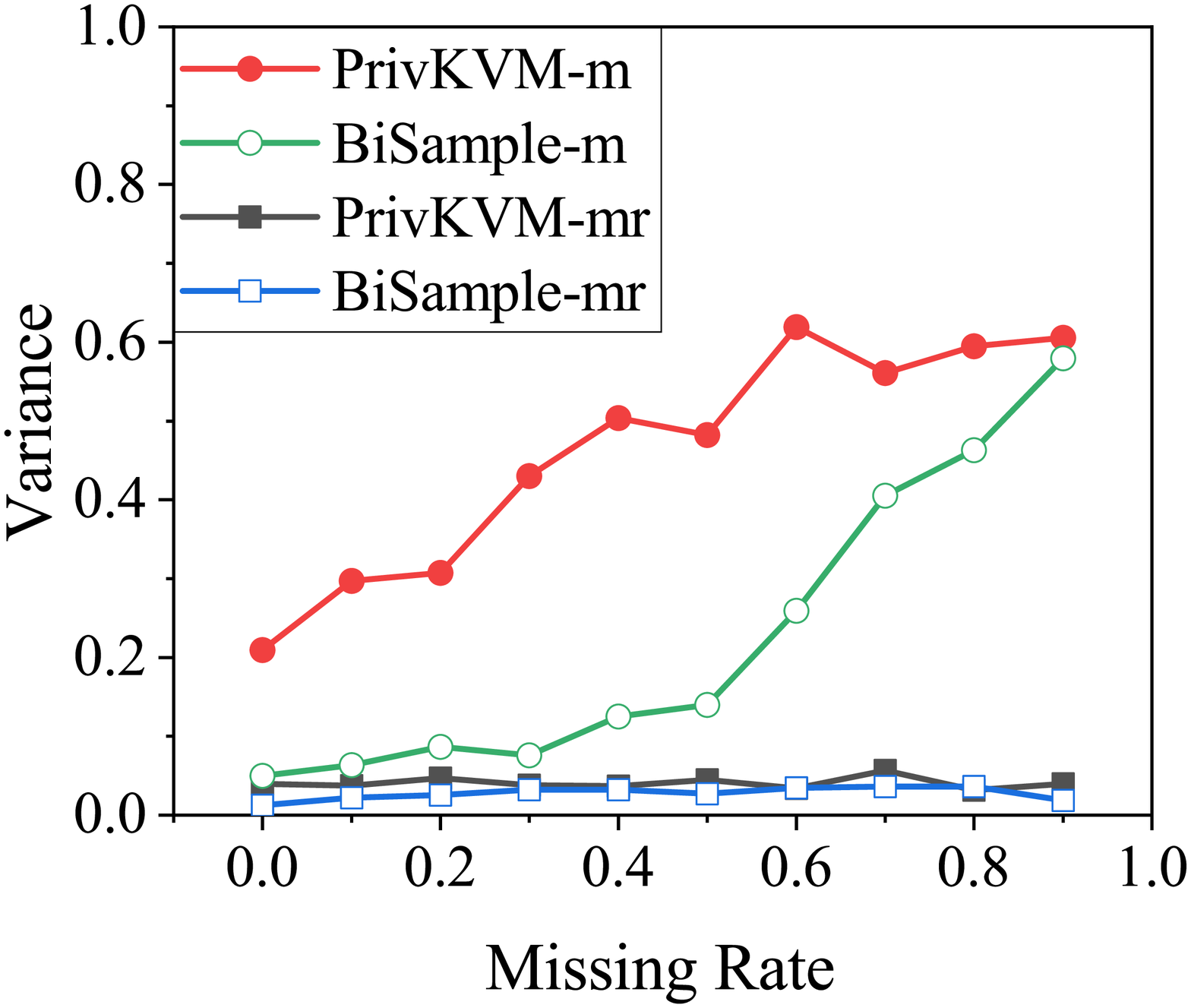} & 
			\includegraphics[width=0.512\textwidth]{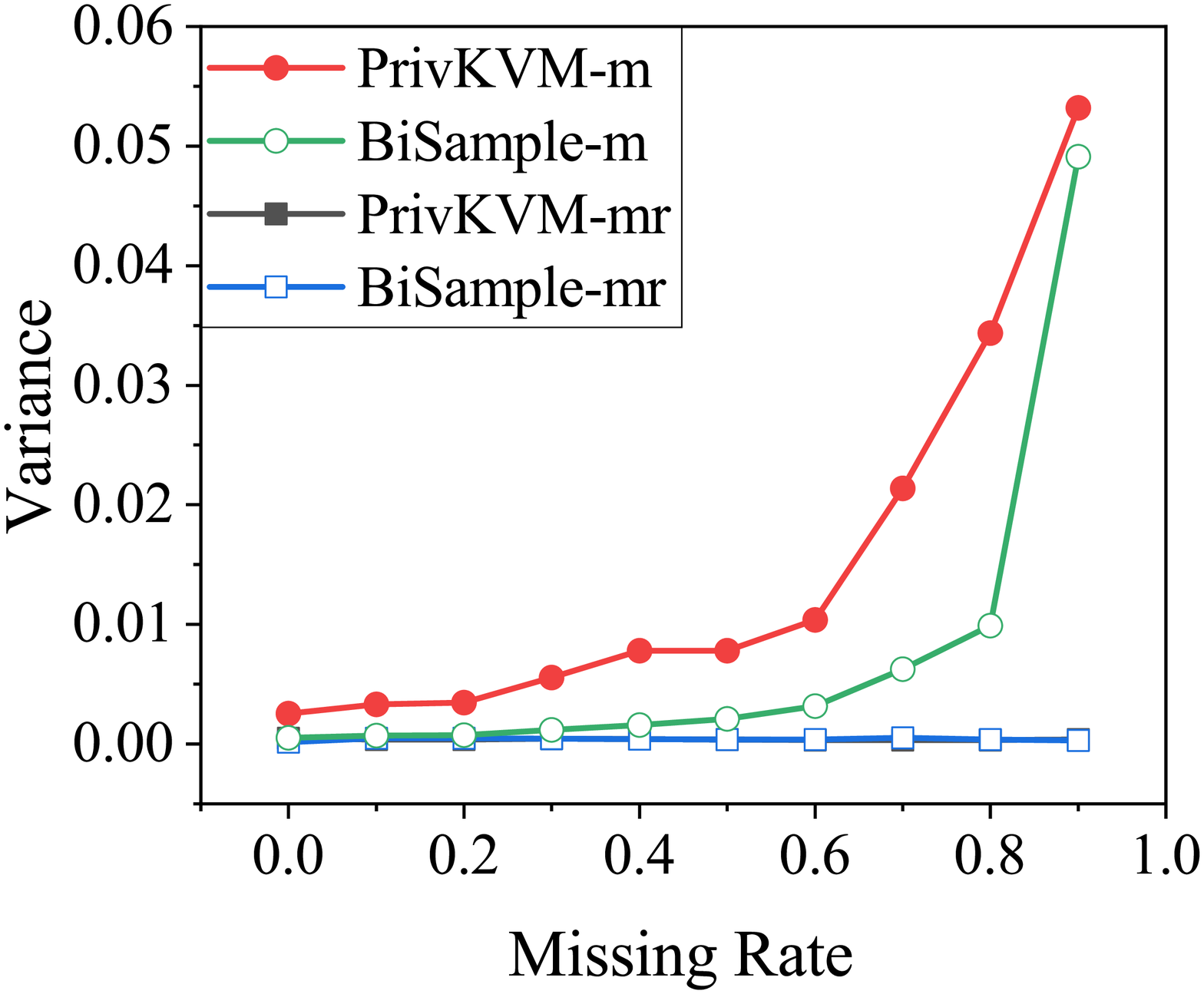}  \\
			(c) GAUSS ($\epsilon=0.1$). &
			(d) GAUSS ($\epsilon=1$).
		\end{tabular}
		\caption{Variance Missing Rate.}
		\label{fig: missing rate changes}
	\end{figure}
	
	The prior experiments are based on the assumption that users' privacy preferences follow a Gaussian distribution. The main reason why users' privacy preferences greatly impact the estimation error is that the rate of users who provide the real value for perturbation changes. Thus in this part of experiment, we directly explore the influence of the missing rate. We fix $\epsilon \in \{0.1, 1\}$ and vary the missing rate to evaluate the estimation error. We only compare the BiSample with PrivKVM as they both can be used for missing rate estimation.
	
	The results are shown in Fig.~\ref{fig: missing rate changes}. As expected, for all of the approaches, the utility measurements decrease when the privacy budget increases. When the missing rate is too high, the mean estimation becomes meaningless, as few data can be used. We also observe that for both missing rate estimation (the curve with ``-mr'') and mean estimation (the curve with ``-m''), the proposed BiSample is superior to PrivKVM. We think the reason is that compared with PrivKVM, we only use value for sampling, thus the introduced noise is lower to that of PrivKVM. Another observation is that in each experiment, the missing rate estimation is more accurate than the mean estimation. The reason is that the missing rate estimation only uses typical randomized response, while the perturbation process involves both discretization and perturbation. 

\subsection{Varying Size of Data}
	
	\begin{figure}[t]
		\centering
		\footnotesize
		\begin{tabular}{cc}
			\includegraphics[width=0.5\textwidth]{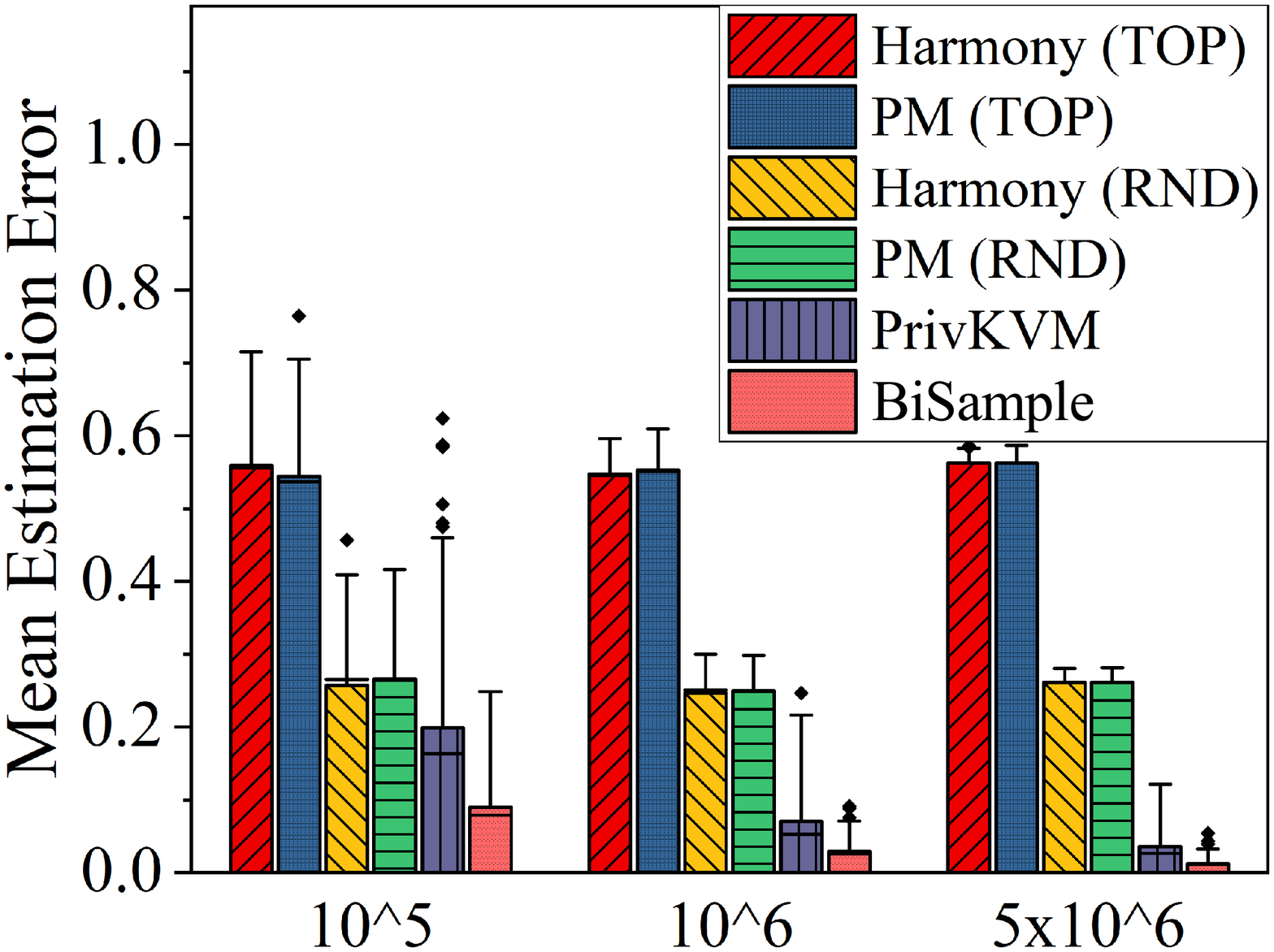} & 
			\includegraphics[width=0.5\textwidth]{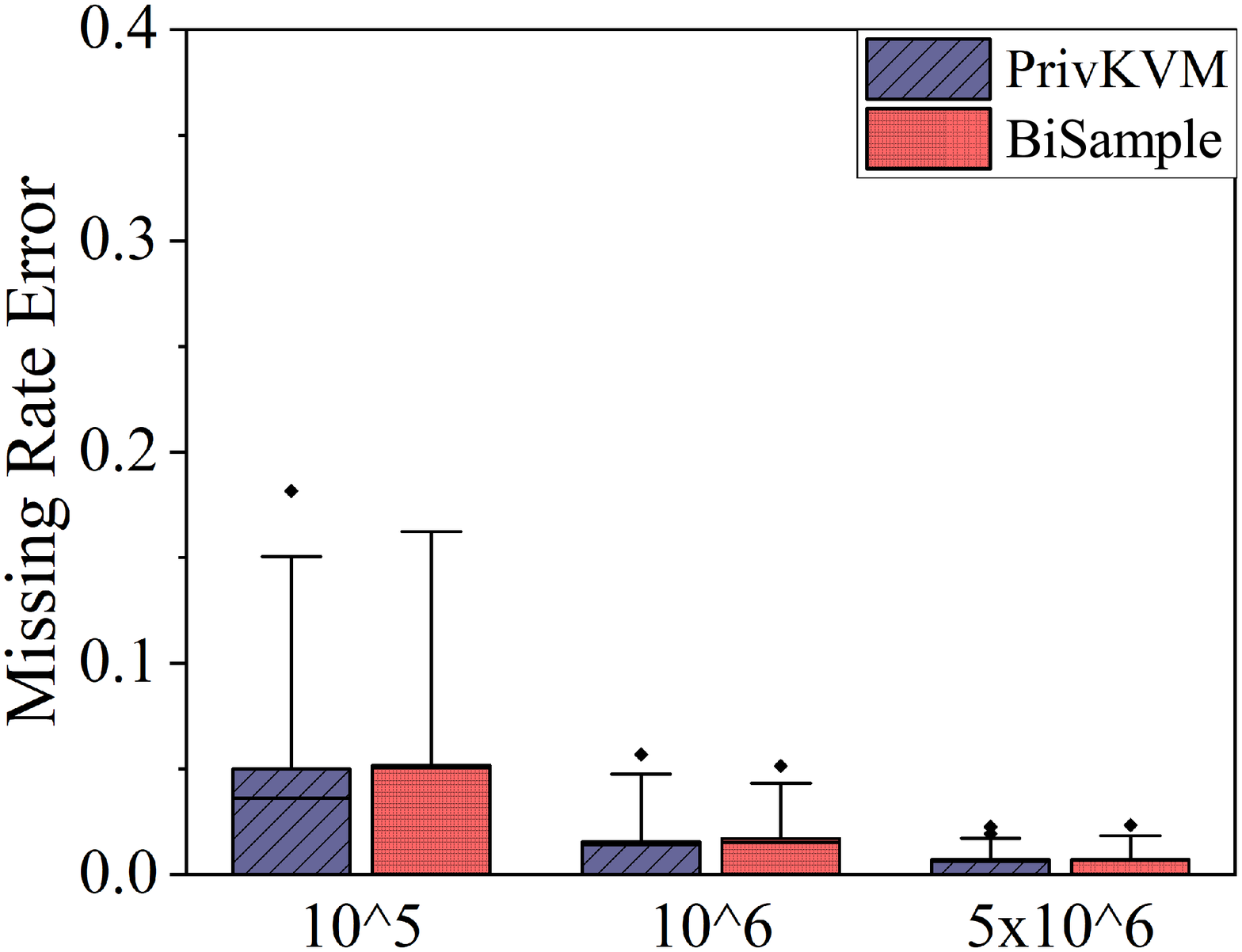}  \\
			(a) Mean Estimation. ($\epsilon=0.1$)&
			(b) Missing Rate Estimation. ($\epsilon=0.1$)
		\end{tabular}
		\caption{Estimation Performance varying Data Size.}
		\label{fig: cmp_data_size}
	\end{figure}
	
	We also consider the influence of the size of data on mean estimation and missing rate estimation. We use the absolute error for evaluation. In the mean estimation of Fig.~\ref{fig: cmp_data_size}(a), we first observe that for PrivKVM and BiSample, the estimation error decreases with a larger size of data. The error of both Harmony and PM is very high because the main inaccuracy is that these two mechanism can not handle the missing data. We observe from Fig.~\ref{fig: cmp_data_size}(b) that the difference between PrivKVM and BiSample is not obvious in terms of missing rate estimation. Overall, the BiSample outperforms PrivKVM in terms of mean estimation. 

\section{Conclusion}
\label{sec: conclusion}

	In this paper, we research the influence of users' privacy-preserving preferences on mean estimation in the framework of LDP. We first propose BiSample, a bidirectional sampling technique for value perturbation. Then users' privacy preferences are considered to avoid fake answers from the user side. Experimental results show that the proposed mechanism can be used for both conventional mean estimation and null-value perturbation with LDP guarantees.

\end{document}